\documentclass[twocolumn, amsmath,nobibnotes,nofootinbib,showpacs,aps,prb]{revtex4}  
\usepackage[dvips]{graphicx,color}
\begin{document}
\title{Glassy magnetic phase driven by short range charge and magnetic ordering in nanocrystalline La$_{1/3}$Sr$_{2/3}$FeO$_{3-\delta}$: Magnetization, M\"{o}ssbauer, and polarised neutron studies}
\author{Sk. Sabyasachi$^1$}
\author{M. Patra$^1$}
\author{S. Majumdar$^1$}
\author{S. Giri$^1$}
\affiliation{$^1$Department of Solid State Physics, Indian Association for the Cultivation of Science, Jadavpur, Kolkata 700032, India}
\author{S. Das$^2$}
\author{V. S. Amaral$^2$}
\affiliation{$^2$Department of Physics and CICECO, University of Aveiro, 3810-193,
Aveiro, Portugal}
\author{O. Iglesias$^3$}
\affiliation{$^3$Departament de F\'{i}sica Fonamental and Institute of Nanoscience and Nanotechnology (IN2UB), Facultat de F\'{i}sica, Universitat de Barcelona, Av. Diagonal 647, 08028 Barcelona, Spain}
\author{W. Borghols$^4$} 
\affiliation{$^4$JCNS, Forschungszentrum J\"ulich, Outstation at FRMII, Lichtenbergstrasse 1, 85747 Garching, Germany}
\author{T. Chatterji$^5$}
\affiliation{$^5$Science Division, Institut Laue-Langevin BP 156, 38042 Grenoble Cedex 9, France}

\begin{abstract}
The charge ordered La$_{1/3}$Sr$_{2/3}$FeO$_{3-\delta}$ (LSFO) in bulk and nanocrystalline forms are investigated using ac and dc magnetization, M\"{o}ssbauer, and polarised neutron studies. A complex scenario of short range charge and magnetic ordering is realized from the polarised neutron studies in nanocrystalline specimen. This short range ordering does not involve any change in spin state and modification in the charge disproportion between Fe$^{3+}$ and Fe$^{5+}$ compared to bulk counterpart as evident in the M\"{o}ssbauer results. The refinement of magnetic diffraction peaks provides magnetic moments of Fe$^{3+}$ and Fe$^{5+}$ are about 3.15$\mu_B$ and 1.57$\mu_B$ for bulk, and 2.7$\mu_B$ and 0.53$\mu_B$ for nanocrystalline specimen, respectively. The destabilization of charge ordering leads to magnetic phase separation, giving rise to the robust exchange bias (EB) effect. Strikingly, EB field at 5 K attains a value as high as 4.4 kOe for average size $\sim$ 70 nm, which is zero for the bulk counterpart. A strong frequency dependence of ac susceptibility reveals cluster-glass like transition around $\sim$ 65 K, below which EB appears. Overall results propose that finite size effect directs the complex glassy magnetic behavior driven by unconventional short range charge and magnetic ordering, and magnetic phase separation appears in nanocrystalline LSFO.
\end{abstract}
\maketitle
\section{Introduction}
Charge ordering (CO) is involved with a phase transition occurring mostly in strongly correlated materials. As a result of strong interaction, charge is localized on different sites in a regular pattern which was first discovered in magnetite.\cite{verwey} Over the last decades charge ordering phenomenon in hole-doped manganites has been investigated extensively, because this intriguing phenomenon is closely related to the delicate interplay between spin, charge, orbital, and lattice degrees of freedom.\cite{rao1} Recent reviews on CO propose that CO is accompanied by the symmetry breaking, resulting in a rich consequence of ferroelectricity which also leads to the multiferroicity having tremendous technological applications for future generation of memory devices.\cite{brink,rao2} This stimulates a renewed attention for understanding the phenomenon in various systems. The CO state can be perturbed by several external parameters. The CO phenomenon can collapse due to application of magnetic field, external pressure or chemical pressure and usually ferromagnetism appears.\cite{Raveau1} Recently, CO phenomenon has been investigated in nanocrystalline  manganites.\cite{Bhat,Idas,Jirac,Zhang,Sama} It has been shown that size reduction induces a weakening of CO and the appearance of weak ferromagnetism as a consequence of size reduction. The particle-size driven intricate interplay between antiferromagnetic (AFM) CO, ferromagnetic (FM), and reentrant-spin-glass-like states was observed in  Nd$_{0.8}$Na$_{0.2}$MnO$_3$.\cite{Sama} A surface phase separation in nanocrystalline charge ordered compounds has been proposed by Dong {\it et al}., leading to the exchange bias (EB) effect based on a phenomenological model.\cite{Dong1} Surface spin-glass (SG) state has been proposed in nanocrystalline Sm$_{0.5}$Ca$_{0.5}$MnO$_3$\cite{Nath} and Nd$_{0.5}$Ca$_{0.5}$MnO$_3$\cite{Liu} to interpret the observed EB phenomenology. Appearance of surface ferromagnetism resulting from size effect and EB at the FM and AFM interface were proposed in nanocrystalline Pr$_{0.5}$Ca$_{0.5}$MnO$_3$.\cite{Zhang1} In fact, exchange bias phenomenology in terms of core/shell model in magnetic nanoparticles has been recently reviewed in detail by Iglesias {\it et al}. \cite{iglesias} These intriguing results inspire to investigate on other nanocrystalline CO compounds apart from extensively studied nanocrystalline manganites.   

Recently, CO phenomenon have been thoroughly investigated by Park {\it et al} in entire series of  $R_{1/3}$Sr$_{2/3}$FeO$_3$ ($R$ = La, Pr, Nd, Sm, and Gd).\cite{Park} The compound with composition La$_{1/3}$Sr$_{2/3}$FeO$_{3-\delta}$ (LSFO) exhibited least rhombohedral lattice distortion at the CO phase transition, $T_{CO}$ = 198 K which is accompanied by an AFM spin ordering.\cite{Kawa} Neutron diffraction studies revealed that CO was gradually developed below 200 K with a charge disproportionation, 2Fe$^{4+} \Rightarrow$ Fe$^{3+}$ + Fe$^{5+}$ in a sequence of Fe$^{5+}$Fe$^{3+}$Fe$^{3+}$Fe$^{5+}$Fe$^{3+}$Fe$^{3+}$... along body diagonal [111] direction with respect to ideal perovskite structure.\cite{Maq,Battle,Yang} The FM exchange between Fe$^{3+}$-Fe$^{5+}$ pairs ($J_F$) and AFM exchange between Fe$^{3+}$-Fe$^{3+}$ pairs ($J_{AF}$) were proposed by McQueeney \textit{ et al.} to fulfill interesting scenario of CO driven by magnetic interactions for $\left| J_F / J_{AF}\right| >$ 1.\cite{Maq} Measurements of resonant soft x-ray magnetic scattering revealed anomalous quasi-2$D$ ordering of 3$d$ spins and 2$p$ holes in LSFO, although this compound has been recognized as a 3$D$ lattice system.\cite{oka} Recently, grain size dependent studies were performed in the 80-200 nm range of LSFO where systematic increase of FM component resulting from the decrease in grain size was demonstrated even at room temperature.\cite{Gao} The weak ferromagnetism was conjectured from appearance of coercivity at room temperature which was suggested to be correlated with the lattice distortion, in which volume of the unit cell increases with decreasing grain size. The destabilization of CO and appearance of ferromagnetism due to reduction of grain size are rather typical manifestation of charge ordered compounds, \cite{Bhat,Idas,Jirac,Zhang,Gao} although elucidation of  microscopic origin of emerging ferromagnetism has been less probed so far.\cite{Jirac,Tapati,Dhital} This can be established through careful investigations by means of microscopic experimental tools such as neutron, NMR, M\"{o}ssbauer studies. 

In this study, we report appearance of glassy magnetic phase driven by unusual short range CO and magnetic ordering in nanocrystalline LSFO which is realized from polarised neutron studies and frequency dependent ac susceptibility measurements. A robust EB effect is observed confirming the magnetic phase separation in nanocrystalline specimen with average size $\sim$ 70 nm. The EB is absent for the bulk counterpart and even absent for nanocrystalline specimen with average size $\sim$ 200 nm. 

\section{Experimental details}
Bulk polycrystalline and nanocrystalline specimens with composition La$_{1/3}$Sr$_{2/3}$FeO$_{3-\delta}$ were prepared by standard solid state reaction and sol-gel technique,\cite{De} respectively. Preheated La$_2$O$_3$ at 1000$^{\circ}$C, SrCO$_3$, and Fe$_2$O$_3$ were used as starting materials. Proper amount of citric acid was used as precursor for synthesizing nanocrystalline specimen. The precursor powders were calcined in the range 1000$^{\circ}$-1200$^{\circ}$C. Addition structural phases appeared when precursor was heated below 1000$^{\circ}$C. For polycrystalline specimen final heating was done at 1400$^{\circ}$C. To achieve desired oxygen stoichiometry both the polycrystalline and nanocrystalline specimens were annealed at 1000$^{\circ}$C in an atmospheric pressure of oxygen. X-ray powder diffraction pattern was recorded in a SEIFERT X-ray diffractometer (Model: XRAY3000P) using Cu K$\alpha$ radiation. Rietveld refinement of diffraction pattern ensures absence of secondary phase in the samples. Grain sizes and grain interior crystalline states were investigated using a field emission Scanning Electron Microscope (FESEM) of model: JSM-6700F and a high-resolution  Transmission Electron Microscope (HRTEM)  of Model: JEOL, 2010. ac and dc magnetometry were carried out in a commercial SQUID magnetometer of Quantum Design (MPMS, Evercool) and vibrating sample magnetometer (VSM) of Cryogenics, UK. $^{57}$Fe M\"{o}ssbauer spectra were recorded in a transmission geometry using a 25 mCi $^{57}$Co source in a Rh matrix with a velocity drive unit of Fast Comtec GmbH in a constant acceleration mode which was coupled with a closed-cycle cryogenics (JANIS) fitted to a vibration-free isolation stand. All the hyperfine parameters obtained from the fits are estimated with respect to the values of metallic $\alpha$-Fe.

Neutron diffraction experiment was carried out on polycrystalline and nanocrystalline LSFO using polarised diffractometer DNS at the FRMII (Garching, Germany). The neutron wave length was 4.74 \AA. We placed samples in an Al foil that was wrapped in to a hollow cylindrical shape. We then put the wrapped samples in side an Al container in the He atmosphere. We recorded diffraction intensities at selective temperatures in the range 3 - 300 K. The diffracted intensity could be separated by the polarization analysis into following three 
contributions: (1) coherent nuclear scattering, (2) spin-incoherent nuclear scattering, and (3) magnetic scattering.\cite{Sch} The refinement of the magnetic diffraction data is done using FULLPROF refinement program.

%
\begin{figure}[t]
\centering
\caption {(Color online) X-ray powder diffraction patterns (black symbols) at 300 K for (a) Bulk and (b) Nano specimen of LSFO. Solid  curve is the Rietveld fit. The lowermost plot is the residual. The bars show the peak positions. (c) HRTEM image of a particle of nanocrystalline specimen. (d) SEM image of the particles. Inset of (d) displays size distribution satisfying log-normal distribution function.} 
\label{XRD}
\end{figure}

\section{Experimental results}
\subsection{Structural properties}
X-ray powder diffraction studies are performed on nanocrystalline specimens finally heated in the range 1000$^{\circ}$-1200$^{\circ}$C. Two representative examples of X-ray diffraction patterns  are displayed in Figs. \ref{XRD}(a) and \ref{XRD}(b) for the specimens annealed at 1000$^{\circ}$C prepared from sol-gel route and 1400$^{\circ}$C prepared from solid state reaction, respectively. To simplify our discussion, henceforth, we address 'Nano' and 'Bulk' for specimens annealed at 1000$^{\circ}$ and 1400$^{\circ}$C, respectively. Average grain size of Nano and Bulk are $\sim$ 70 nm and $\sim$ 1 $\mu$m, respectively as confirmed from FESEM images. Fits of the powder diffraction patterns at 300 K using Rietveld refinement technique are shown in Figs. \ref{XRD}(a) and \ref{XRD}(b). The difference plot at the bottom confirms absence of secondary phase for both the cases. The results reveal rhombohedral crystal structure (space group: $R\overline{3}c$) in the hexagonal setting.\cite{Battle,Yang} The refined parameters are summarized in Table \ref{Str}. We note that lattice parameters of the bulk specimen are consistent with the reported results.\cite{Battle,Yang} The refined parameters for Nano given in Table \ref{Str} show a minor structural change compared to the bulk counterpart. Unit cell volume slightly increases up to 0.12 \% for Nano specimen. This increase is rather very small compared to previous observation ($\approx$ 3 \%) in nanocrystalline LSFO with average size 80 nm.\cite{Gao} This marked difference may be correlated to the crystallinity of LSFO nanoparticles. As evident in Fig. 1(c) grain interior plane extents until the edge of grain boundary where good crystallinity causes less lattice distortion. The Fe-O1-Fe bond angle is reduced to 172.9$^{\circ}$ for Nano from 177.4$^{\circ}$ for Bulk. The larger deviation from 180$^{\circ}$ for Nano indicates appearance of additional structural disorder with reducing grain size.  

\begin{table} 
\centering
\caption{Structural parameters of nanocrystalline (Nano) and polycrystalline (Bulk)  LSFO from the analysis of X-ray Diffraction Data at 300 K using Rietveld refinement. The atomic sites of La/Sr and Fe are at (0,0,0.25) and (0,0,0), respectively.}
\label{Str}
\begin{center}
\begin{tabular}{ccc}		
\hline
\hline
LSFO            &            Bulk         & Nano      \\
\hline
space group     &            $R\bar{3}c$       &      $R\bar{3}c$        \\
\multicolumn{3}{c}{\bf{Lattice parameters}}  \\
$a$ (\AA) 		& 				 5.4767(2)								&     5.4775(2)	\\  
$c$ (\AA) 		& 				13.3946(6)									&   13.4063(7)		\\
volume (\AA$^3$)	&	    347.93					        &			 348.34		\\
\multicolumn{3}{c}{\bf{Atomic positions}}  \\

O1 ($x$)			&					0.4920(6)									  	&		0.4782(3)		  \\
O1 ($y$)			&				    0										  &			  0		   \\
O1 ($z$)			&					0.25										&		   0.25		  \\
\multicolumn{3}{c}{\bf{Reliability factors}}  \\
$R_p$ (\%)		& 				2.0576											&		   2.9485			  \\
$R_{wp}$ (\%) &				1.7563											&		  1.8328		  \\
$\chi^2$			&				1.1715										&		   1.6087	  \\			
\multicolumn{3}{c}{\bf{Bond angles and Bond distances}}  \\	
Fe-O1-Fe ($^\circ$)	        & 				177.4						&		172.9		      \\
La/Sr-O1$ (\AA)$					  & 		2.7360	&		 2.740		\\
Fe-O1$ (\AA)$		& 		1.938	&		 1.940	  \\
\hline
\end{tabular}
\end{center}
\end{table} 

As depicted in Fig. \ref{XRD}(c) intraplaner spacing (0.25 nm) of HRTEM image is close to the calculated distance of (110) plane estimated from X-ray diffraction pattern. We carefully note that any secondary phase is not found at the grain boundary of HRTEM image. The FESEM image is displayed in Fig. \ref{XRD}(d). Inset shows the bar diagram of  particle size distribution which could be fitted with log-normal distribution function with a mean value $\sim$ 70 nm and standard deviation = 0.21. We further note from the FESEM image (not shown here) that average size of the particles is $\sim$ 200 nm when final heating was done at 1100$^{\circ}$C.

\begin{figure}[t]
\centering
\includegraphics[width = \columnwidth]{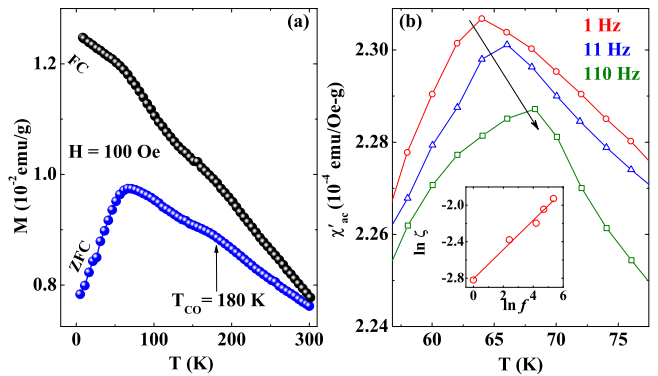}
\caption {(Color online) (a) Thermal variation of ZFC-FC magnetization measured at 100 Oe for Nano specimen. (b) Real part of ac susceptibility [$\chi_{ac}^{\prime}(T)$] displayed at selective $f$ = 1, 11, and 110 Hz and  $H_{ac}$ = 4 Oe. Inset exhibits the fit using dynamical scaling law.} 
\label{T-ac-dc}
\end{figure}

\subsection{Magnetization results and exchange bias effect}
Thermal variation of ZFC-FC (zero-field cooled and field-cooled) magnetization recorded in 100 Oe is displayed in Fig. 2(a) for Nano specimen. The FC magnetization was recorded in the warming cycle. The ZFC and FC magnetization do not meet each other at 300 K which is much above paramagnetic to AFM ($T_N$) and CO ($T_{CO}$) ordering temperature. The result is consistent with the appearance of coercivity at 300 K for nanocrystalline LSFO reported by Gao {\it et al}.\cite{Gao} This probably happens due to survival of magnetic  ordering until 300 K in the Nano specimen, which is further confirmed in the current study by polarised neutron results. Both in ZFC and FC magnetization of Nano specimen a diffused signature of $T_{CO}$ is observed around 180 K [indicated by the arrow in Fig. 2(a)], which is well below $T_{CO}$ at 198 K for the Bulk specimen. At low temperature another well defined maximum appears around $\sim$ 60 K in the ZFC magnetization which was not convincingly observed in the bulk counterpart.\cite{Park} This maximum emerges to be a glassy magnetic transition as evident from the frequency dependent ac susceptibility ($\chi_{ac}$) results. A broadened signature in the FC curve is also observed, below which a steady increase is noticed with decreasing temperature. 

The $\chi_{ac}$ is measured at frequency, $f$ = 1, 11, 66, 110, and 211 Hz with ac field, $H_{ac}$ = 4 Oe. Thermal variation of real part of $\chi_{ac}$ [$\chi_{ac}^{\prime}(T)$] around the maximum observed in ZFC magnetization is displayed in Fig. 2(b) at selected frequencies. A strong $f$-dependent peak shift is noticed, which could be fitted with the dynamical scaling law close to phase transition at $T_f$. The scaling law relates the critical relaxation time, $\tau_{max}$ to the correlation length ($\zeta$) as $\tau_{max}=\tau_0\zeta^{z\nu}$, where $\zeta$=$T_0/(T_f-T_0)$, $\tau_0$ is the microscopic flipping time, $z$ is the dynamic exponent, $\nu$ is another exponent related to spin-correlation length, and $T_0$ provides the value of $T_f$ at $f$ $\rightarrow$ 0. The best fit is shown in the inset of Fig. 2(b). The value of $T_0$ obtained is 60.6 K which is close to the maximum obtained from dc magnetization. The fit further provides $z\nu \approx$ 6, which holds good in the range between 4 and 12, typically found for atomic SG compounds. The value of $\tau_0 \approx 4 \times 10^{-8}$ s is much slower than the values in the range, $\sim 10^{-12} - 10^{-14}$ for atomic SG compounds. \cite{binder} The values of $\tau_0$ are typically found to be slower for the cluster-glass (CG) and nanocrystalline compounds than the values for atomic SG compounds. The values of $\tau_0$ are $\sim 10^{-10}$ s for CG La$_{0.95}$Sr$_{0.05}$CoO$_3$,\cite{nam} $\sim 10^{-10}$ s for CG La$_{1-\delta}$Mn$_{0.7}$Fe$_{0.3}$O$_3$,\cite{de1} $\sim 10^{-10}$ s for nanocrystalline  La$_{0.88}$Sr$_{0.12}$CoO$_3$,\cite{patra1} $\sim 10^{-7} - 10^{-9}$ s for nanocrystalline Co$_{50}$Ni$_{50}$ alloy.\cite{thakur1} Current investigation on dynamic ac susceptibility measurements confirm a disordered glassy magnetic transition at $T_f$.

\begin{figure}[t]
\centering
\includegraphics[width = \columnwidth]{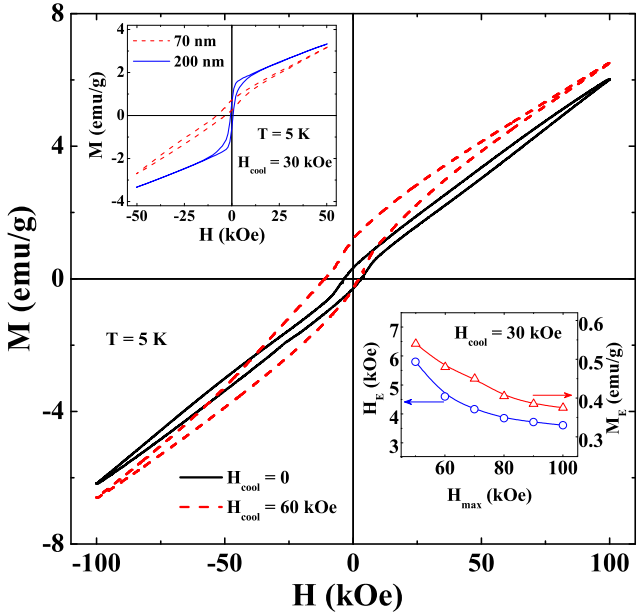}
\caption {(Color online) (a) The MH loop at 5 K measured in between $\pm$ 100 kOe after cooling in ZFC and FC ($H_{\rm cool}$ = 60 kOe) modes for Nano specimen. Upper inset shows absence of shift for 200 nm size compared to large shift for 70 nm size due to field cooling at 5 K. Lower inset shows plots of $H_E$ and $M_E$ shifts with $H_{max}$ at 5 K for $H_{cool}$ = 30 kOe.} 
\label{M-H}
\end{figure}

Magnetic hysteresis (MH) loops were measured at various temperatures after cooling the sample in ZFC and FC (in various cooling fields, $H_{cool}$) modes. At 5 K representative examples of MH loop after cooling in ZFC and FC ($H_{cool}$ = 60 kOe) modes are displayed in Fig. 3. Magnetization does not show any saturating trend for measurements up to $\pm$ 100 kOe. First, we note that while the ZFC loop is symmetric, in the FC mode,  the symmetry is lost and a shift of the loop along both axes is clearly observed. Moreover, a huge enhancement (nearly double) of coercivity is noticed due to field cooling. These are typical manifestations of EB effect.\cite{nogues1,giri} We note that EB effect manifested by the loop shift is absent for Bulk specimen. This is even absent for nanocrystalline LSFO with average grain size $\sim$ 200 nm. As seen in the upper inset of Fig. 3 shift is absent for LSFO with 200 nm average size compared to large shift with 70 nm average size when measurement is carried out at 5 K within $\pm$ 50 kOe after cooling in $H_{cool}$ = 30 kOe. 

Horizontal and vertical shifts are defined as EB field ($H_E$) and EB magnetization ($M_E$), respectively. The $H_E$ is determined from  shift in the $H-$axis at $M$ = 0 and $M_E$ is determined from the vertical shift at 100 kOe.\cite{nogues1,nogues2,giri} Substantial values of $H_E$ = 4.4 kOe and $M_E$ = 0.5 emu/g are observed at 5 K for $H_{cool}$ = 60 kOe. The proper choice of maximum field ($H_{max}$) applied for recording a MH loop  is crucial for obtaining $H_E$ and $M_E$, because small $H_{max}$ may lead to minor loop effects.\cite{patra2,nogues3,oscar} The plots of $H_E$ and $M_E$ with $H_{max}$ are shown in the lower inset of Fig. 3. We note that the increasing and decreasing field branches of MH loops recorded up to $\left|H_{max}\right|$ smaller than 50 kOe, do not join at the maximum applied field.  Therefore, the values of $H_{E}$ and $M_{E}$ are given in the plot for 50 kOe $\leq \left|H_{max}\right| \leq$ 100 kOe, where 100 kOe is the highest achievable field of our VSM facility. Both plots show that $H_E$ and $M_E$ decrease rapidly with increasing $\left|H_{max}\right|$ and approach toward stabilized values at $\left|H_{max}\right|$ = 100 kOe. Although current value of $H_E$ is smaller than the highest reported value ($\approx$ 8 kOe at 4.2 K for $H_{cool}$ = 100 kOe) for Nd$_{60}$Fe$_{30}$Al$_{10}$, \cite{Hong} this value is however, substantially large among the reported values in structurally single phase alloys and compounds.\cite{giri} 

\begin{figure}[t]
\centering
\includegraphics[width = \columnwidth]{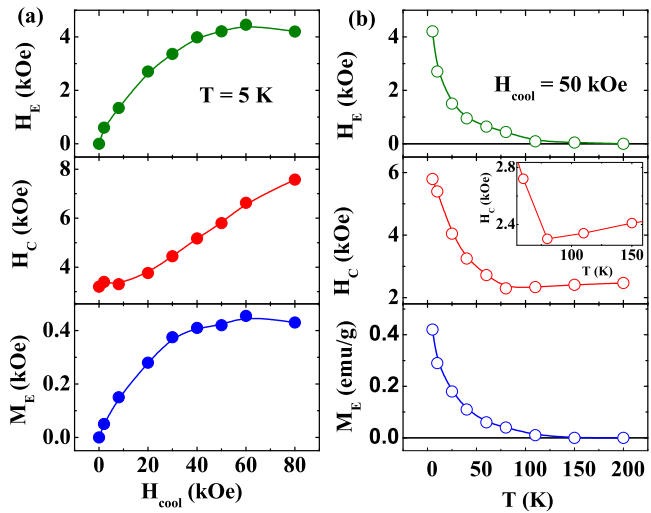}
\caption {(Color online) (a) At 5 K $H_{cool}$ dependence of $H_E$ (top panel), $H_C$ (middle panel), and $M_E$ (bottom panel) for Nano specimen. (b) Thermal variation of of $H_E$ (top panel), $H_C$ (middle panel), and $M_E$ (bottom panel) for $H_{cool}$ = 50 kOe for Nano specimen.} 
\label{EB}
\end{figure}

The dependence of $H_E$, $H_C$, and $M_E$ on $H_{cool}$ is shown in Fig. 4(a). Both $H_E$ and $M_E$ exhibit a similar dependence on $H_{cool}$. A sharp increase is observed with increasing $H_{cool}$ up to 60 kOe, above which it shows a saturating trend. The increase of $H_E$ and $M_E$ is accompanied by an increase of $H_C$ that does not seem to saturate for $H_{cool}$ up to 80 kOe. This observation is an indication of appearance of a magnetic phase with substantial increased anisotropy induced by the cooling field. The behavior can be understood by noticing that the field cooling protocol induces the development of a new layer composed of pinned spins at the interface between two magnetic phases that causes the observed EB effect. \cite{giri}  Anisotropy of this cooling field driven layer composed of pinned spins strongly depends on the individual anisotropy of the magnetic phases coupled with this pinned layer. We note that the $H_{cool}$ dependence of $H_C$ in the current investigation is distinctly different from the observation in a classical combination of FM/AFM Co/CoO nanostructures where the $H_{C}-H_{cool}$ plot displayed a saturating trend above 10 kOe.\cite{SDas} Although AFM component is the main component in the current investigation, the $H_{C}-H_{cool}$ plot indicates strong anisotropy of the pinned layers. The effect of randomness in the spin alignment at this layer averages $H_E$ to zero at small $H_{cool}$ but, with increasing $H_{cool}$, the pinned spins of the hard magnetic phase progressively align into the field cooling direction giving rise to the appearance of an increasing $H_E$ as well as $H_C$.  In the current investigation this progressive alignment of spins driven by increased cooling field is not completed even at 80 kOe and at 5 K. 

Thermal dependencies of $H_E$, $H_C$, and $M_E$ are displayed in Fig. 4(b). A substantial decrease of both $H_E$ and $M_E$ with increasing $T$ is observed and both quantities vanish close to 100 K. The $H_C$ shows a similar trend up to $\sim$ 80 K that changes toward an increase for higher temperatures [see Inset of middle panel of Fig.4(b)]. This change in behavior indicates a change in overall anisotropy around $\sim$ 80 K. This is analogous to that addressed by Nogu\'{e}s {\it et al}. in a layered system composed of FM and AFM substances.\cite{nogues1} It was pointed out that anisotropy of the AFM component changes with $T$ which resulted in the significant change in the $H_{C}-T$ plot.

\begin{figure}[t]
\centering
\includegraphics[width = 0.7\columnwidth]{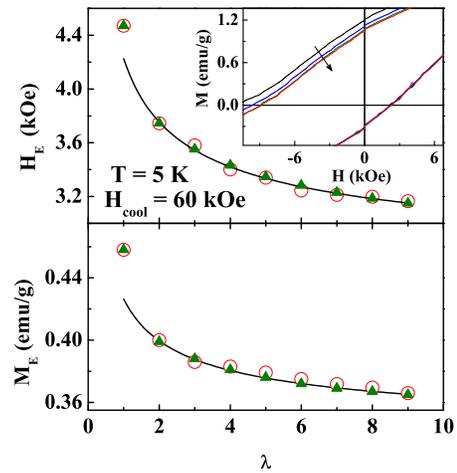}
\caption {(Color online) Dependence of $H_E$ (top panel) and $M_E$ (bottom panel) with number of field cycling, $\lambda$ (indicated by open symbols) at 5 K after cooling the sample in $H_{cool}$ = 60 kOe for Nano specimen. Continuous curves reveal the fit using the empirical formula whereas solid symbols shows the calculated data from the recursive formula described in the text. Inset of top panel displays magnified MH loop, displaying training effect. Direction of arrow indicates curves with increasing field cycling.} 
\label{TE}
\end{figure}

The training effect (TE) describes the systematic decrease of the loop shift due to successive field cycling after cooling the sample in a static magnetic field and it is commonly found in systems, displaying EB effect.\cite{giri} The signature of TE is clearly demonstrated in the inset of top panel of Fig. 5, wherein horizontal shift decreases with increasing number of field cycling ($\lambda$). The plots of $H_E$ and $M_E$ vs. $\lambda$ displayed in top and bottom panels of Fig. 5, respectively exhibit a decrease with $\lambda$. This decrease can be fitted to an empirical formula, $H_E - H_E^{\infty} \propto \lambda^{-1/2}$ for $\lambda \geq$ 2, where $H_E^{\infty}$ is the value of $H_E$ at $\lambda = \infty$.\cite{giri} Since this formula cannot describe initial sharp decrease of $H_E$ and $M_E$, a generalized interpretation of TE was proposed by Binek which is described by the recursive formula $H_E (\lambda + 1) - H_E (\lambda) = -\gamma [H_E (\lambda) - H_E (\lambda = \infty)]^3$,\cite{binek} where $\gamma$ is a sample dependent constant. Both formulae describe correctly the $H_E$ and $M_E$ vs. $\lambda$ plots for all $\lambda$ as seen in Fig. 5 (continuous line and solid symbols). The values of  $H_E(\lambda = \infty)$ and $M_E(\lambda = \infty)$ obtained from the fit are 2.44 kOe ($\gamma$ = 0.086 kOe$^{-2}$) and 0.323 emu/g [$\gamma$ = 23.28 (emu/g)$^{-2}$], respectively. These values at $\lambda = \infty$ are substantial, which reinforces the fact that the observed shifts are genuinely EB effects and do not emerge due to minor loop effects.

%
\begin{figure}[t]
\centering
\includegraphics[width = 0.9\columnwidth]{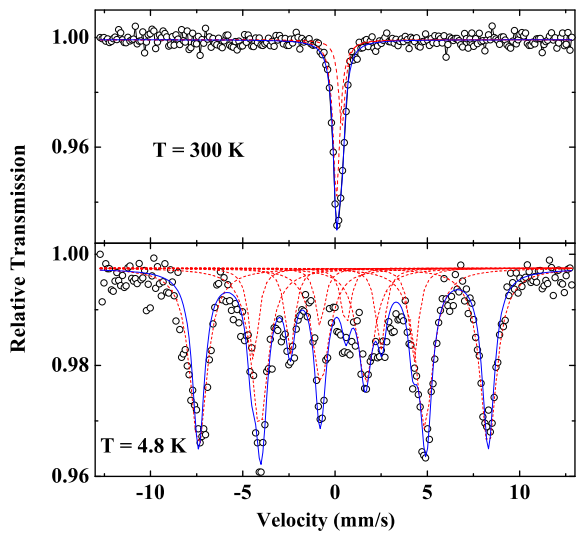}
\caption {(Color online) M\"{o}ssbauer spectra recorded at 300 and 4.8 K for Nano specimen. Continuous curves display least square fit of the spectra while broken curves demonstrate the individual components.} 
\label{Mossbauer}
\end{figure}

\begin{table*} 
\centering
\caption{Hyperfine field ($B_{hf}$), Isomer shift (IS), Quadrupole Splitting (QS), and relative Intensity (Int) at 300 and 4.8 K as obtained from the fits of the M\"{o}ssbauer Spectra.} 
\label{Tbl}
\begin{center}
\begin{tabular}{ccccccccccccc}		
\hline
&  \multicolumn{4}{c}{Fe$^{3+}$} & &   \multicolumn{2}{c}{Fe$^{4+}$} & & \multicolumn{4}{c}{Fe$^{5+}$}  \\
\hline
T    & 	B$_{hf}$	 &IS	   &QS   & Int&       &IS          & Int.&    &B$_{hf}$	  &IS     &QS   &Int\\
 (K)     & (kOe)        &(mm/s)   &       &  (\%) &             &(mm/s)           &(\%) &         & (kOe)       &(mm/s)   &        &(\%)\\
\hline
 300(nano)     & $-$      &0.39$\pm$0.01   & $-$     &34$\pm$0.5  &             & 0.09$\pm$0.01     & 66$\pm$0.5   &       &$-$          &$-$           & $-$       &$-$ \\
 300(Bulk)$^*$     & $-$        &0.258       &$-$       &35    &          &0.064      &65      &    & $-$            & $-$       &$-$       & $-$\\
\hline
5 (nano)       &488$\pm$0.5    &0.47$\pm$0.01     &-0.11$\pm$0.01   &67.5$\pm$0.5    &     &$-$          & $-$           &      &270$\pm$0.5     &-0.13$\pm$0.01     &-0.025$\pm$0.005     &32.5 $\pm$0.5 \\
20 (bulk)$^*$       &474       &0.393     &-0.036  &67  &        & $-$        &$-$          &      &266       & -0.035    & -0.016     &33\\
\hline
\noindent$^*$ Reference [18] 
\end{tabular}
\end{center}
\end{table*}

\subsection{M\"{o}ssbauer results}
To justify EB effect and ratio between Fe$^{3+}$ and Fe$^{5+}$ in Nano specimen, M\"{o}ssbauer study is performed both in Nano and Bulk specimens. M\"{o}ssbauer spectra were recorded at 300 and 4.8 K for Nano specimen which is displayed in Fig. 6. At 300 K a singlet spectrum is observed over nearly smooth background. This indicates non-existence of any secondary phase ascribed to iron content. The spectrum composed of two components is evident from the fit shown in the top panel of Fig. 6. The isomer shift, IS = 0.396 mm/s corresponds to Fe$^{3+}$ and the rest signifies for Fe$^{4+}$. As shown in Table II, the values of IS and intensity distribution obtained for Nano specimen are in accordance with that observed for the polycrystalline compound.\cite{Yang} A magnetically split sextet spectrum is revealed at 4.8 K which is also fitted into two sextet components. The values of hyperfine parameters obtained from the fit of spectrum at 4.8 K  are compared in Table II with the reported polycrystalline values measured at 20 K.\cite{Yang} We note that hyperfine parameters are reasonably close to the values obtained for polycrystalline compound. This confirms the coexistence of Fe$^{3+}$ and Fe$^{5+}$ components as found in the bulk counterpart. As reported for polycrystalline compounds appearance of Fe$^{5+}$ and increase of intensity of Fe$^{3+}$ at 4.8 K occur due to charge disproportion, 2Fe$^{4+} \Rightarrow$ Fe$^{3+}$ + Fe$^{5+}$. The value of intensity ratio between Fe$^{3+}$ and Fe$^{5+}$ is 2.07:1, which is close to desired ratio (2:1), signifying oxygen stoichiometry close to desired value at  $\delta$ = 0 for Nano specimen.

%
\begin{figure}[t]
\centering
\includegraphics[width = 0.8\columnwidth]{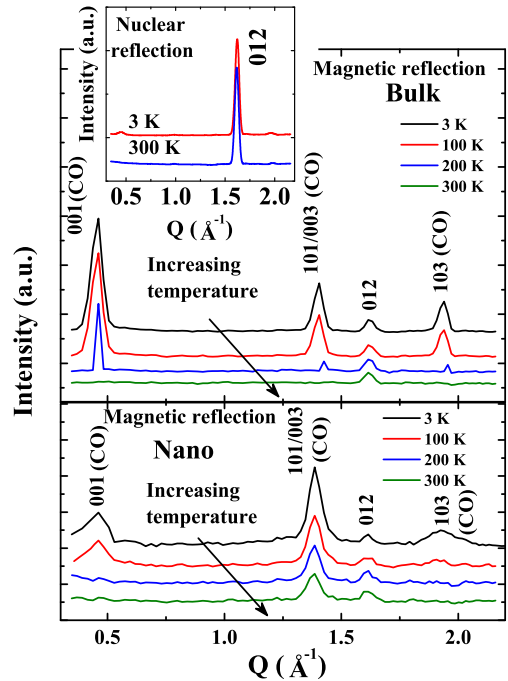}
\caption {(Color online) Polarised magnetic reflections for Bulk (top panel) and Nano (bottom panel) specimens at selective temperatures. Inset of top panel display diffraction diagram of nuclear channel at 3 and 300 K for Bulk.  Indexing of the planes are displayed in the figure.} 
\label{Neutron}
\end{figure}

\subsection{Polarised neutron results}
To probe nature of CO and magnetic ordering in Nano specimen exhibiting substantial EB, polarised neutron measurements are carried out on both the Nano and Bulk specimens. The magnetic reflections of Nano and Bulk specimens are displayed in Fig. 7 at selective temperatures, in which intensities are plotted as a function of $Q = 4\pi\sin\theta/\lambda$. Indexing of magnetic reflections are done according to the rhombohedral structure with $R\bar{3}c$ space group. The polarised magnetic reflections at 3 K for Bulk are in accordance with the magnetic diffraction data at 15 K reported by Yang {\it et al}.\cite{Yang} As seen in Fig. 7 an additional weak peak (012) is observed in the the polarised magnetic reflections for both the Bulk and Nano specimens which was absent in the previous reports. It is to be noted that in case of strong nuclear intensity this signature can appear as a weak peak in spin flip channel because of deviation from the perfect polarization situation. Inset of the top panel of Fig. 7 displays a very strong intense diffraction peak at (012) plane in the diffraction diagram of nuclear channel from 3 to 300 K. 

Below $T_{CO}$ three CO peaks at (001), (101)/(003), and (103) planes appear for Bulk. At 200 K these three peaks just appear and intensities of the peaks increase considerably at 100 K. These intensities almost saturate below 100 K as evident by nearly same intensities of these peaks at 3 and 100 K. This indicates completion of long range CO and magnetic ordering processes at 100 K for Bulk specimen. According to Yang {\it et al}. lowest-$Q$ peak corresponding to (001) plane signifies degree of Fe$^{3+}$ and Fe$^{5+}$ charge ordering as well as AFM ordering and the rest two CO peaks corresponding to (101)/(003) and (103) planes arise from AFM ordering. Although magnetic reflections of Bulk specimen reproduce the reported magnetic diffraction results for Bulk, intensity patterns of these magnetic reflections observed at same $Q$ and thermal variation of each reflection are significantly different for Nano. 

The CO peaks at (001) and (103) planes are broadened with much reduced intensities. We note that these two peaks are not visible at 200 K unlike results for  Bulk. This is in accordance with the ZFC magnetization results where $T_{CO}$ shifts toward lower temperature at 180 K for Nano specimen. Magnetic reflections at (001) and (103) planes are evident at 100 K and intensities increase with decreasing temperature unlike our observation for Bulk. This indicates short range charge and magnetic ordering processes and are not settled at 100 K for Nano. The considerable broadening with reduced intensities indicates that short range charge and magnetic ordering involve wide distribution of grain size in the range $\sim$ 40-100 nm [as seen in the inset of Fig. 1(d)]. 

We note that widths of the magnetic reflections are much broader than the resolution of the instrument and also considerably broader than the particle size broadening of the diffraction peaks. In such a case average magnetic coherence length has been obtained\cite{tc,wang} from the line width broadening of the diffraction peak \textcolor{red}{using Scherrer formula\cite{patt}. The calculation provides} $\sim$ 25 nm which is significantly smaller than that of the average physical size of the nanoparticles.\cite{tc} The result is significantly different from the reported magnetic coherence length which is nearly same in magnitude with the physical size.\cite{see} This indicates that surface of the particles is in disordered magnetic state which does not take part in the magnetic ordering. The disordered magnetic spins at the surface leads to the glassy magnetic behavior as evident in the $f$-dependent $\chi^{\prime}_{ac}(T)$ results. Unlike variation of intensities of (001) and (103) planes, intensity of magnetic reflection at other CO (101)/(003) plane increases monotonically with decreasing temperature. In fact, this reflection is evident much above $T_{CO}$ even at 300 K for Nano which is absent for  Bulk. This is in accordance with that observed ZFC-FC effect of magnetization for Nano specimen. As seen in Fig. 2(a) the ZFC-FC curves do not meet at 300 K. It is also pertaining to point out that Gao {\it et al}. reported appearance of coercivity in the magnetization curve at 300 K for nanocrystalline LSFO.\cite{Gao} The polarised neutron results thus confirm survival of short range ordering, which leads to the appearance of coercivity as well as considerable difference between ZFC and FC magnetizations at 300 K.

%
\begin{figure}[t]
\centering
\includegraphics[width = 0.8\columnwidth]{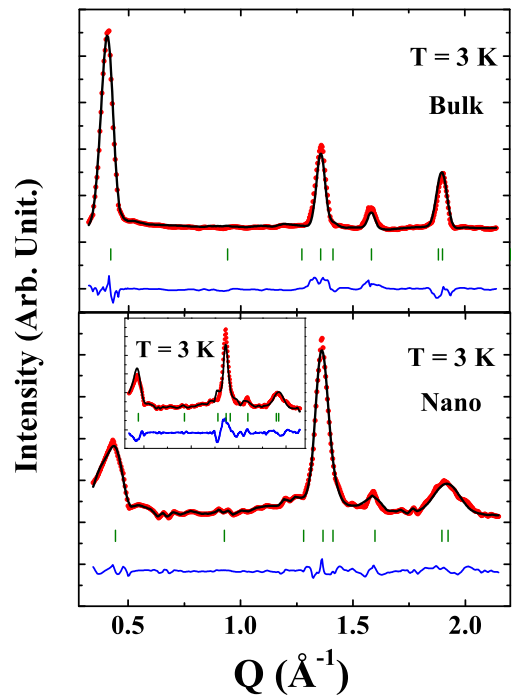}
\caption { (Color online) Rietveld refinement of the polarised magnetic diffraction data for Bulk (top panel) and Nano (bottom panel) specimens at 3 K. \textcolor{red}{In the bottom panel inset shows the refinement using modified model as described in the text.}} 
\label{mag-fit}
\end{figure}

 The magnetic diffraction data for Bulk and Nano are fitted in the $P$1 space group using Rietveld refinement done previously for polycrystalline LSFO.\cite{Battle,Yang} The unit cell of the magnetic structure is considered same as that of the crystalline structure. This includes six iron atoms at  positions (0, 0, 0), (1/3, 2/3, 1/4), (2/3, 1/3, 1/3), (0, 0, 1/2), (1/3, 2/3, 2/3) and (2/3, 1/3, 3/4) 
with a charge sequence of charge ordering, Fe$^{5+}$Fe$^{3+}$Fe$^{3+}$Fe$^{5+}$Fe$^{3+}$Fe$^{3+}$.... In the refinement magnetic moments of the iron atoms are restricted as two groups (Fe$^{5+}$ and Fe$^{3+}$) to form an antiferromagnetic structure where absolute values of the magnetic moments are maintained the
same for each group. The magnetic moments entirely lie in the basal plane. In accordance with the proposed spin configurations by Yang {\it et al}.\cite{Yang} the satisfactory fit considering Fe$^{5+}(\uparrow)$Fe$^{3+}(\uparrow)$Fe$^{3+}(\downarrow)$Fe$^{5+}(\downarrow)$Fe$^{3+}(\downarrow)$Fe$^{3+}(\uparrow)$ spin configuration is displayed in the top panel of Fig. 8, where bars show the position of diffraction peaks and difference plot is shown at the bottom. The quite satisfactory fit provides the refined moments for Fe$^{3+}$ and Fe$^{5+}$ to be about 3.15$\mu_B$ and 1.57$\mu_B$, respectively at 3 K \textcolor{red}{with reliability factors, $\chi^2$=3.32 and $R_{mag}$=13.89\%.} The values are in accordance with the reported values about 3.0$\mu_B$ and 1.3$\mu_B$, respectively for Fe$^{3+}$ and Fe$^{5+}$ at 15 K.\cite{Yang}

Although the intensity profile of Nano specimen is much weaker than the Bulk counterpart, a similar refinement is carried out on magnetic diffraction data of Nano specimen at 3 K considering the same charge ordering sequence. This assumption is reasonable, because analysis of the M\"{o}ssbauer spectrum at 4.8 K points to nearly the same ratio between Fe$^{3+}$ and Fe$^{5+}$ in Nano compared to Bulk specimen. The quite satisfactory fit is displayed at the bottom panel of Fig. 8 \textcolor{red}{with reliability factors,  $\chi^2$=1.10 and $R_{mag}$=15.47\%, that are reasonable compared to the previous \cite{Battle} and more  recent reports.\cite{melot,chmai}} The fit provides the refined moments for Fe$^{3+}$ and Fe$^{5+}$ to be about 2.7$\mu_B$ and 0.53$\mu_B$, respectively at 3 K. 

\textcolor{red}{To test authenticity, a slightly modified model is tried for the refinement with a modified sequence where nearest neighboring Fe$^{3+}$ and Fe$^{5+}$ sequences are exchanged. The best refinement provides $\chi^2$=2.25 and $R_{mag}$=46.60\%, although the moment values are nearly the same as obtained using the Bulk model. To test further, the refinement is done with another modified sequence by exchanging alternate Fe$^{3+}$ and Fe$^{5+}$ sequences. This also provides nearly same values of moment with $\chi^2$=7.05 and $R_{mag}$=39.29\%. The mismatch of the fitted intensity profile with the experimental data  is evident in the inset of the bottom panel of Fig. 8 for example, in case of first modified sequence. Thus,  the results justify better reliability of the refinement for Nano with model justified for Bulk and we conclude that} the values of the refined moments for Nano are significantly smaller than that obtained for Bulk specimen. \textcolor{red}{The reduced moment may appear due to readjustment of spin structure through spin canting for Nano. In this case, spins are homogeneous throughout the particles. If ferromagnetic component emerges due to spin canting, additional magnetic contribution on the top of nuclear peak may appear  for Nano which is not observed in Fig. 7. The core-shell structure is the other possibility which has been frequently proposed for magnetic nanoparticles. The core acts like Bulk and the shell behaves like glassy magnetic component consisting of disordered magnetic spins. The scattering from the core only contribute to the magnetic Bragg intensities and magnetic scattering from the glassy magnetic region goes to the diffuse background which is not taken into account in the refinement process. However, the refinement leads to the reduction of refined magnetic moment as recently evident for MnO nanoparticles.\cite{fey,wang} Similar core-shell magnetic structure has recently been proposed for MnO nanoparticles where considerably reduced moment was reported based on the refinement of the magnetic diffraction data for MnO nanoparticles.} The much reduced Mn-moment compared to the bulk counterpart was suggested to be due to the appearance of 20\% to 80\% disordered surface spins depending on the size of the nanoparticles.

\begin{figure}[t]
\centering
\includegraphics[width = 0.7\columnwidth]{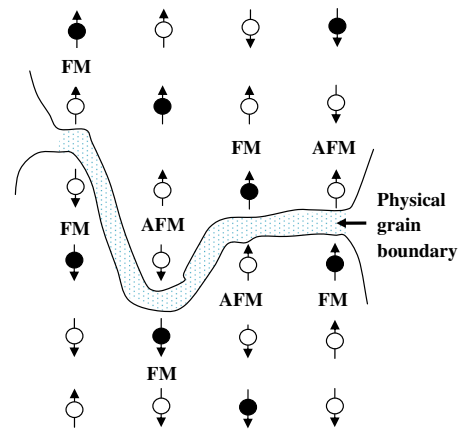}
\caption {Schematic representation of charge and magnetic ordering of Fe$^{5+}$ (solid circles) and Fe$^{3+}$ (open circles) spins in the (001) plane. Possible physical boundary and super-exchange interactions at the grain boundary are highlighted.} 
\label{Cartoon}
\end{figure}

\section{Discussions and conclusions}

M\"{o}ssbauer and polarised neutron results elucidate vital issues of CO and magnetic ordering	of Nano specimen, which could not be probed through bulk magnetization studies. Neutron results clearly demonstrate short range CO and magnetic ordering. This short range ordering process is not uniform for all planes. Although CO between Fe$^{3+}$ and Fe$^{5+}$ has been tailored in Nano specimen, analysis of M\"{o}ssbauer spectrum confirms that spin state as well as ratio of Fe$^{3+}$ and Fe$^{5+}$ do not alter compared to bulk counterpart. To interpret robust EB effect, these crucial results facilitate to propose a phenomenological model displaying possible phase separation scenario in Nano specimen. Figure 9 displays a schematic representation of random possible discontinuity in long range CO as well as super-exchange (SE) paths in the (001) plane due to finite size effect. Solid circles in the figure stand for Fe$^{5+}$ whereas open circles correspond to Fe$^{3+}$. Position of oxygen between metal atoms for the linkage of SE paths is not shown here for simplicity. As seen in the figure Fe$^{5+}$-O-Fe$^{3+}$ SE is FM while Fe$^{3+}$-O-Fe$^{3+}$ SE interaction is AFM according to our proposed model of spin configuration.\cite{Maq} Because of discontinuity in the SE paths ascribed to finite size effect a random possible array of FM and AFM SE interactions are illustrated at the grain boundary (GB). Emergence of competing FM and AFM SE interactions lead to the spin frustration at the GB. This frustration along with the disorder attributed to various sources viz., distribution of grain size, defects at GB, various possible array of FM and AFM SE interactions, may lead to glassy magnetic phase in the magnetic nanoparticles. Possible surface SG state has been conjectured for nanocrystalline Sm$_{0.5}$Ca$_{0.5}$MnO$_3$\cite{Nath} based on $f$-dependent $\chi_{ac}(T)$ results and for  Nd$_{0.5}$Ca$_{0.5}$MnO$_3$ from the bulk magnetization study.\cite{Liu} Current $f$-dependent $\chi_{ac}(T)$ results reveal glassy magnetic state, which has been verified in various nanocrystalline alloys and oxides.\cite{patra1,thakur1,bedanta} This glassy magnetic state is in accordance with the polarised neutron results where much reduced Fe$^{3+}$ and Fe$^{5+}$ moments compared to bulk counterpart confirm emergence of disordered magnetic spins for Nano specimen.

Till date, the issues of size effect on suppression of CO or retaining CO state together with possible phase separation has been discussed scarcely in the literature, although majority of the studies are centered around mixed-valent manganites.\cite{Bhat,Idas,Jirac,Zhang,Sama,Dong1,Nath,Liu,Zhang1} Interestingly, the issues of magnetic phase separation driven by suppression of CO has been argued in the various possible ways to interpret the observed results in similar nanocrystalline CO compounds. For example, possible phase separation between FM and AFM was pointed out in nanocrystalline Pr$_{0.5}$Ca$_{0.5}$MnO$_3$.\cite{Zhang1} On the other hand, emergence of surface SG-like phase was addressed in nanocrystalline Sm$_{0.5}$Ca$_{0.5}$MnO$_3$\cite{Nath} and Nd$_{0.5}$Ca$_{0.5}$MnO$_3$,\cite{Liu} displaying dissimilar results and interpretations were done mainly based on bulk magnetization studies. These interesting results attract the community and the issues need to be concluded from meticulous studies using microscopic experimental tool such as neutron studies. The current results elucidate nature of suppression of CO process due to size effect where polarised neutron diffraction and M\"{o}ssbauer studies together with the magnetization results direct the possible phase separation scenario in nanocrystalline LSFO.

In conclusion, we have reported intriguing glassy magnetic phase driven by the short range charge and magnetic ordering in nanocrystalline LSFO. The polarised neutron and frequency dependent ac susceptibility results confirm the glassy magnetic behavior. The magnetic phase separation between glassy magnetic and antiferromagnetic components leads to the substantial exchange bias effect at low temperature resulting from field cooling process, which is absent for the bulk counterpart. On increasing average grain size surface to bulk ratio decreases which results in considerable decrease of disordered surface spins. For $\sim$ 200 nm average size the reduced disordered surface spins can not take part in pinning mechanism due to field cooling and EB effect does not occur. 

\vspace{0.2in}
\noindent
{\bf Acknowledgment}\\
S.G. acknowledges Council of Scientific \& Industrial Research (CSIR), India (Project No. 03(1167)/10/EMR-II) for financial support and DST Nanoscience unit of IACS, Kolkata for TEM and MPMS facilities. S.S. also thank CSIR, India for the Junior Research Fellowship. S.D acknowledges Foundation of Science and Technology of Portugal for the postdoctoral grant. O.I. acknowledges funding of the Spanish MICINN through Grant project MAT2009-08667 and Integrated Spanish-Portuguese Action under AIB2010PT- 00099, European Union FEDER funds ("Una manera de hacer Europa") and Catalan DIUE through Project  2009SGR856.

\end{document}